# To reduce soil salinity: the role of irrigation and water management in global arid regions across development phases

First author: Haiyang Shi[1,2,3]

1 State Key Laboratory of Desert and Oasis Ecology, Xinjiang Institute of Ecology and Geography, Chinese Academy of Sciences, Urumqi, Xinjiang, 830011, China.

2 University of Chinese Academy of Sciences, 19 (A) Yuquan Road, Beijing, 100049, China.

3 Department of Geography, Ghent University, Ghent 9000, Belgium.

**Abstract**

Reducing soil salinization of croplands with optimized irrigation and water management is essential to achieve land degradation neutralization (LDN). The effectiveness and sustainability of various irrigation and water management measures to reduce basin-scale salinization remain uncertain. Here we use remote sensing to estimate soil salinity of arid croplands from 1984 to 2018. We then use Bayesian network analysis to compare the spatial-temporal response of salinity to water management, including various irrigation and drainage methods, in ten large arid river basins: Nile, Tigris-Euphrates, Indus, Tarim, Amu, Ili, Syr, Junggar, Colorado, and San Joaquin. Managers in basins at more advanced phases of development implemented drip and groundwater irrigation, which effectively controlled salinity by lowering groundwater levels. For the remaining basins where conventional flood irrigation is used, economic development and policies are crucial to establishing a virtuous circle of 'improving irrigation systems, reducing salinity, and increasing agricultural incomes' necessary to achieve LDN.

**Main**

Target 15.3 of the UN Sustainable Development Goals (SDG 15.3) proposes to achieve a world with zero net land degradation, termed land degradation neutrality[1] (LDN), by 2030. LDN is defined[1] as 'a state whereby the amount and quality of land resources necessary to support ecosystem functions and services and enhance food security remain stable or increase within specified temporal and spatial scales and ecosystems.' Reducing soil salinity, which can damage ecosystem health and crop production, is crucial to realizing LDN. Poor irrigation and water management practices—such as excessive irrigation, which can raise groundwater levels—can result in soil salinization, especially in arid areas. Previous estimates suggest that 10% of total arable land (1 billion ha) is affected by soil salinity and that roughly 20% of irrigated lands of the world are salt-affected and essentially commercially unproductive[2,3]. Salinization in irrigated areas of arid regions is closely related to water management and watershed development. Improvements in sustainable development, reductions in agricultural intensification, and improvement in irrigation and drainage methods[4] can be necessary to decrease soil salinity and to achieve LDN. To ensure that salinity control measures can be applied in low-income areas, we must develop long-term and low-cost salinity control measures and assess their effectiveness.

To date, our knowledge about the impact of macro-scale water management strategies on salinization is limited. Most previous studies on this issue are limited to local scales[5–7]. Two factors explain the lack of regional and global-scale understanding: limitations of large-scale spatiotemporal estimates of soil salinity, and difficulties in obtaining accurate macro-scale irrigation and water use data, especially in developing countries in Central Asia and the Middle East, which are heavily affected by salinization. Many studies in recent years have used multispectral, hyperspectral, and microwave satellite imagery data to estimate regional-scale soil salinity based on the relationship between soil salinity and reflectance in specific bands[3,8–10]. Various ancillary variables (e.g., topographic, hydrological, and climatic variables) and machine learning algorithms have also been used to improve the accuracy of estimates[3]. However, the accuracy of global-scale soil salinity estimates remains limited due to the limited sample size of soil salinity data and the high spatial and temporal variability of salinity. These data limitations prevent us from sufficiently understanding soil salinity risks globally and reduce our ability to develop effective strategies to prevent soil salinization and achieve LDN.

Irrigation and water management measures are closely linked to the phase of development within river basins. Generally, as the economy develops, the agricultural development of the river basin may gradually shift to a water-saving, salinity-controlling, and eco-friendly type[11]. Water-saving irrigation is more likely to be used, and more attention paid to the efficiency and sustainability of the agricultural system. Basins at higher phases of watershed development may provide experiences and lessons for basins at the lower phases[12]. This study investigates the relations between soil salinization and irrigation and water management at large watershed scales in ten arid river basins: Nile, Tigris-Euphrates, Indus, Tarim, Amu Darya (Amu), Ili, Syr Darya (Syr), Junggar, Colorado, and San Joaquin (Central Valley). These watersheds vary in economic development and irrigation and drainage practices, which allows us to explore causal relationships between factors related to soil salinity and management practices. This analysis brings a new international perspective to the problem of soil salinization that considers interactions among many factors to inform water management and work to achieve LDN.

**Salinity dynamics derived from the Landsat imagery**

We estimated soil salinity in global arid croplands from 1984 to 2018 using a Random

Forest model[3] on Google Earth Engine (GEE) with Landsat data and in-situ samples combined (Methods). There are significant differences in the fluctuation of the salinity in the ten river basins (Fig. 1). Comparing the two periods of 1984-2000 and 2001-2018, the soil salinity of the Syr, Ili, Tarim, Junggar, and the Nile basins decreased; while the salinity of Indus, Amu, Colorado, and San Joaquin basins was relatively stable; and the salinity of lower Tigris-Euphrates basin has increased significantly. Some basins showed different trends in salinity in different portions of the basins (e.g. increase in salinity in the lower Amu and the mid-stream Syr).

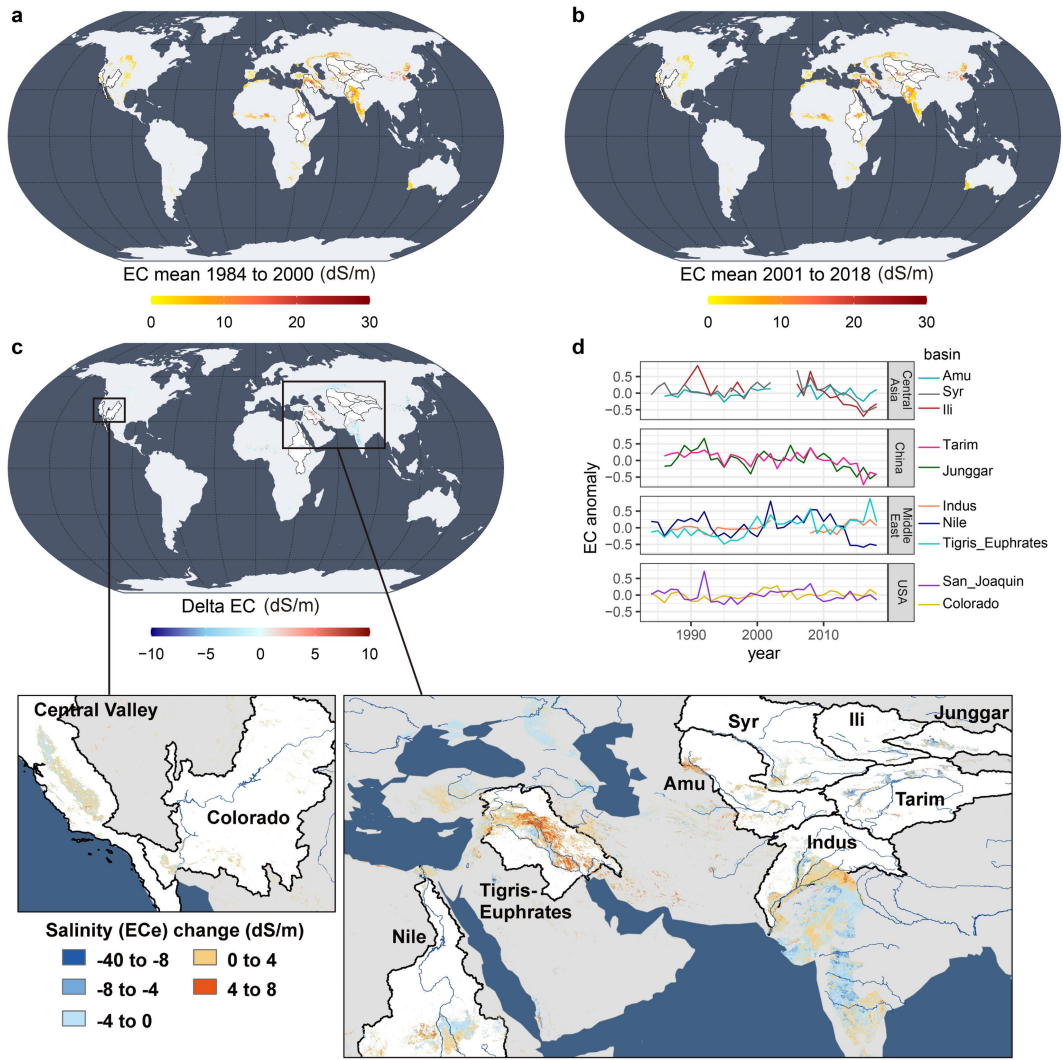

Fig. 1: Spatio-temporal changes in topsoil (0-30cm depth) salinity of cultivated lands in ten river basins in arid and semi-arid regions. **a**, Mean salinity from 1984 to 2000. **b**, Mean salinity from 2001 to 2018. **c**, Salinity change between the period 2001 to 2018 and 1984 to 2000. **d**, Interannual variation (the median value of the salinity z-score) in topsoil salinity of croplands in ten typical river basins: Nile, Tigris-Euphrates, Indus, Tarim, Amu Darya (Amu), Ili, Syr Darya (Syr), Junggar, Colorado and Central Valley. ECe (electrical conductivity EC at the saturated extract phase) is the measure of soil salinity.

## Impacts of irrigation and water management on salinity

These ten basins are currently at different phases of development and differ in the irrigation and drainage methods used (Supplementary Fig. 1) and in investment in salinity control (Fig. 2). In general, most basins have changed from initial small-scale development to large-scale development. Initial development is generally accompanied by large volumes of water being diverted and used in irrigation, followed later by shifts to protective and water-saving methods that are more efficient and sustainable. Salinity management in the Amu, Syr, and Tigris-Euphrates basins is still relatively low[13,14], possibly because of transboundary water disputes and wars (e.g. in Iraq and Syria). The major irrigation systems in Indus and the lower delta of Nile were developed prior to the early 20th century[15,16], but these two basins have not yet widely promoted water-saving micro-irrigation. With the rapid economic development of eastern China, communities in the Tarim and Junggar basins have adopted water-saving irrigation techniques[17] and have become the main cotton planting areas in China. Drip irrigation and plastic-mulching irrigation[18,19] have been widely promoted through improved technologies and government financial support, and have increased yields of agricultural products. Many new croplands were developed with the saved water. The Central Valley and Colorado basins in the western United States used water-saving irrigation earlier due to shortages of water resources, droughts[20], and rising water prices. They increasingly pumped groundwater[7] for irrigation and used tile subdrainage[21].

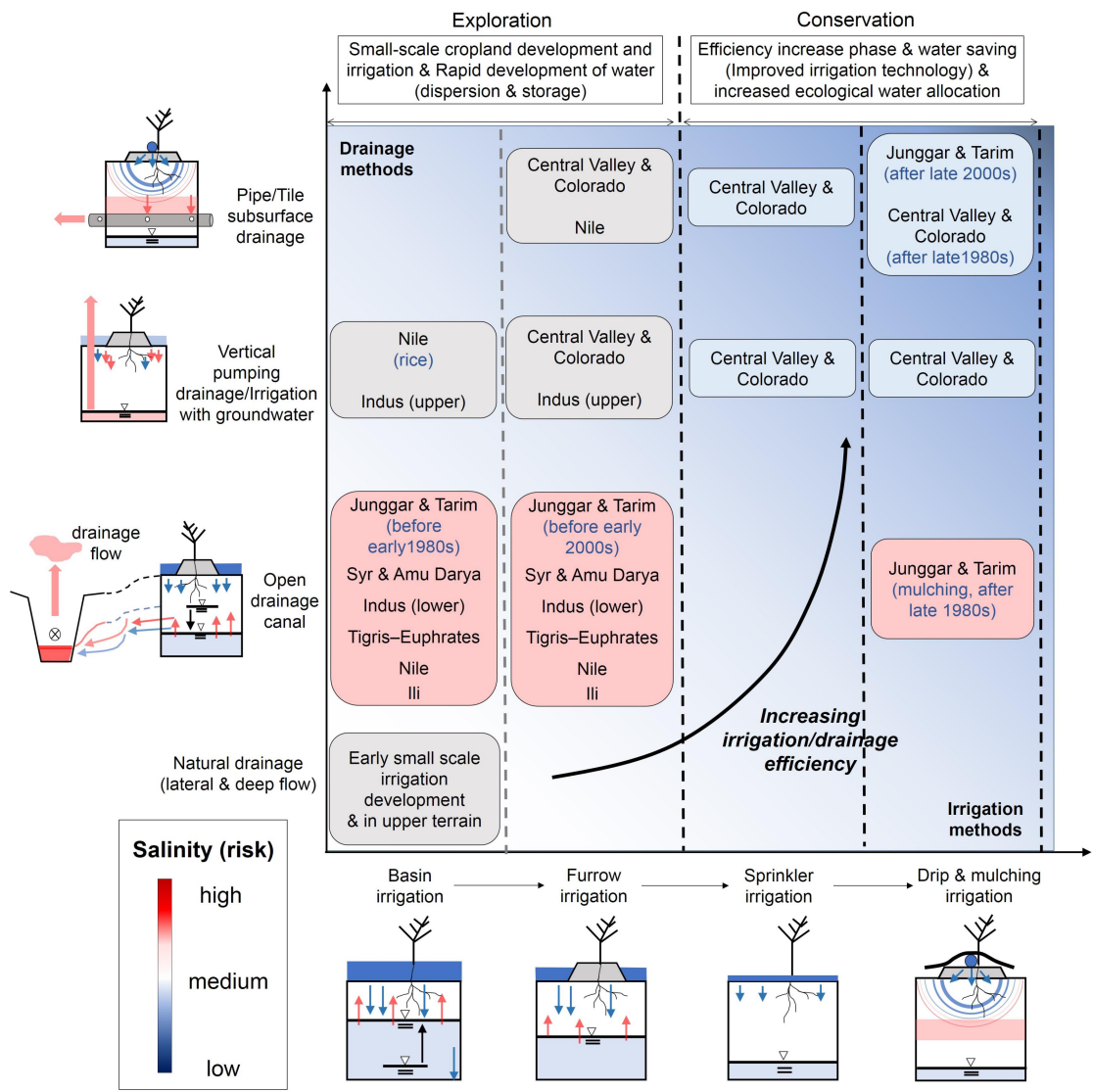

Fig. 2: Irrigation and artificial drainage methods and their impacts on soil salinity in the ten selected river basins at different phases of water resources development. The horizontal axis reflects the evolution of irrigation methods. The vertical axis reflects the evolution of drainage. The arrow reflects the transport of water and salt. Potential groundwater tables are shown in diagrams of irrigation and drainage practices on each axis. Red corresponds to high salinity risk, and blue corresponds to low salinity risk.

We analyzed the linear correlation between water management-related variables and soil salinity at three levels: all-basins combined, individual basins, and individual basins in five-year segments of time (Fig. 3). We also analyzed the linear correlation between water management practices and temporal soil salinity at individual sample sites (Fig. 4). In the Junggar and Tarim basins, drip irrigation fraction is negatively correlated with salinity. In the Junggar Basin, many newly reclaimed croplands have lower soil salinity, while in the lower Tigris-Euphrates Basin, many abandoned croplands in Syria and Iraq have higher salinity (Fig. 4.c). In the upper and mid-stream of the Syr Basin (e.g., the Fergana Valley), basin-scale runoff is negatively correlated with salinity, possibly because more irrigation water is applied in wet years compared to neutral years, which is consistent with the correlation between soil salinity change and runoff change every five years (Supplementary Fig. 2). Lower-layer soil moisture[22] (30 to 150 cm depth of soil, highly correlated with groundwater level and capillary water rise in the global hydrology model PCR-Globwb) is positively correlated with salinity in the alluvial plain below the underground overflow zone in the Junggar Basin, possibly because land reclamation and irrigation reduced the recharge of the surface water to the groundwater and lowered the groundwater level over the last 30 years. The distance to the main drains of Syr and Amu basins is positively correlated with salinity at the basin scale, indicating that the distribution of their drainage canals explains the spatial variation of basin-scale salinity variation (i.e., the main drains are more densely dug and located in areas with higher salinity). In the Tarim and Junggar basins, in which the irrigation system gradually shifted to large-scale drip irrigation, the effectiveness of the drains has decreased. There is a negative correlation between salinity and gross domestic product (GDP) per capita in the Tarim and Junggar basins, possibly caused by the consequent higher investments in irrigation system improvements and salinity control. In contrast, salinity and GDP are positively correlated in basins (e.g. Tigris-Euphrates) where investments in salinity control may not have increased despite increases in GDP per capita. Mulching (data in 2015) is more negatively correlated to salinity in the Junggar Basin from 2014 to 2018 compared to other periods, while tile subdrainage (data in

2017) is negatively correlated to salinity in Colorado from 2014 to 2018.

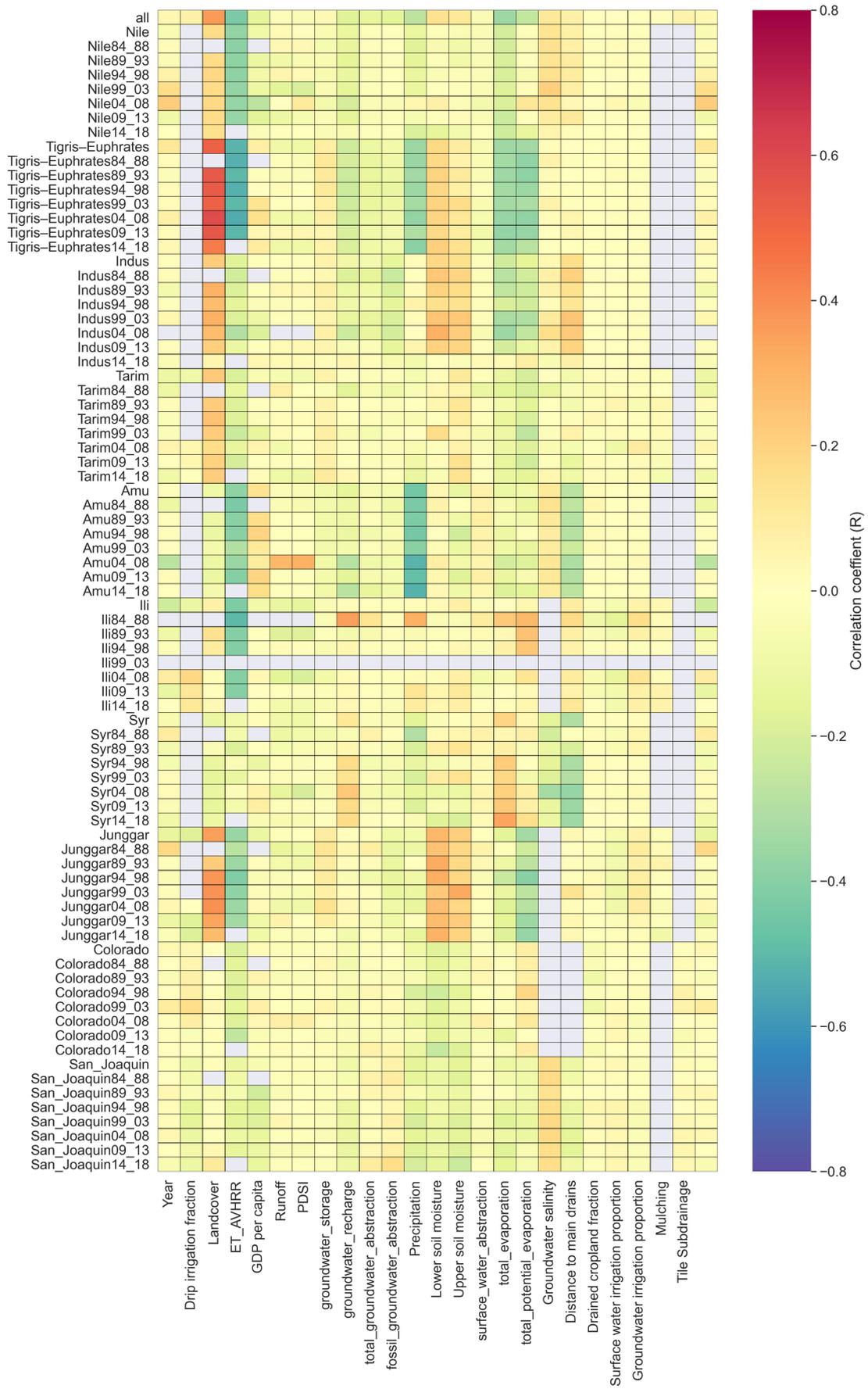

Fig. 3: Linear correlation between Spatio-temporal soil salinity and water management-related variables. Data for the period

from 1984 to 2018 in each basin is divided into five-year subsets, namely 1984 to 1988 (84_88), 1989 to 1993 (89_93), 1994 to 1998 (94_98), 1999 to 2003 (99_03), 2004 to 2008 (04_08), 2009 to 2013 (09_13), and 2014 to 2018 (14_18). 'all' corresponds to the entire data set of the ten basins from 1984 to 2018. 'landcover' was reclassified with value 0 for croplands, and 1 for bare lands and other types for the quantitative correlation analysis. ET_AVHRR indicates actual evapotranspiration data derived from the vegetation index based on AVHRR images (Methods).

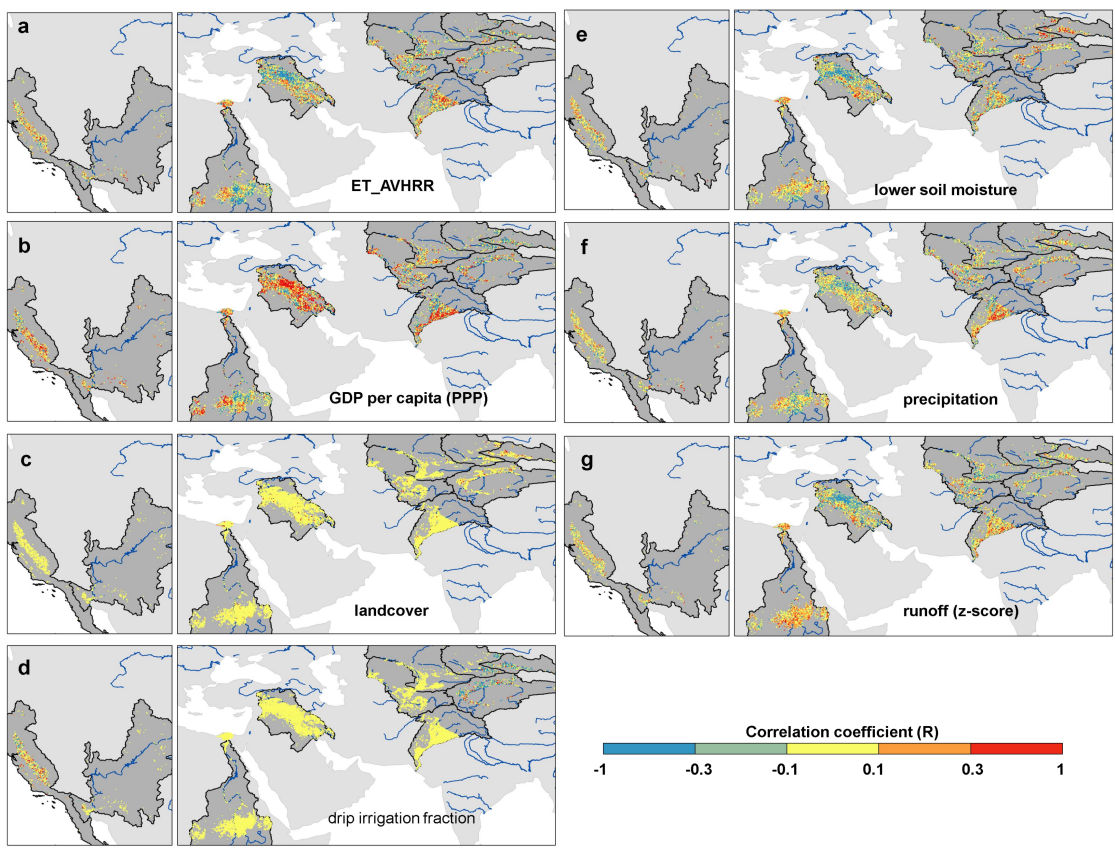

Fig. 4: Linear correlation between temporal soil salinity and water management-related variables. The color of each randomly sampled site (Methods) corresponds to the correlation coefficient between its interannual salinity time series and the time series of covariates. **a, b, c, d, e, f, g** the correlation of salinity time series to (a) ET_AVHRR, (b) GDP per capita, (c) landcover (reclassified with value 0 for croplands, and 1 for bare lands and other types), (d) drip irrigation fraction, (e) lower soil moisture (the 30 to 150 cm depth of soil), (f) precipitation, (g) the z-score of the basin-scale runoff. ET_AVHRR indicates actual evapotranspiration data derived from the vegetation index based on AVHRR images (Methods).

To deepen the understanding of the causal links and remove pseudo-correlations that cannot be explained by process-based perspectives, we used the Bayesian network (BN) to simulate the causal relations between water management and salinity. According to the empirical causality among these variables, we give the causal structure of BN (Fig. 5.a). Using a unified node discretization scheme and subsequent parameterization, we compiled the BN, which includes observations for each basin, and conducted a sensitivity analysis (Methods; Fig. 5.b). Under the constraints of causality assumptions of the BN, precipitation and potential evapotranspiration likely do not cause variation in salinity, although they are strongly correlated to salinity. The sensitivity patterns of salinity to soil moisture and precipitation differ across basins. Instead, irrigation may contribute considerably to variation in salinity. The sensitivity patterns of salinity to Palmer drought severity index (PDSI) and runoff appear to be very similar, indicating that basin-scale drought influences runoff, thereby influencing surface water for irrigation and soil salinity. In the Junggar, Ili, and Tarim basins located in Xinjiang, China, the increase of GDP per capita is causally linked to salinity—increases in GDP lead to increases in drip irrigation.

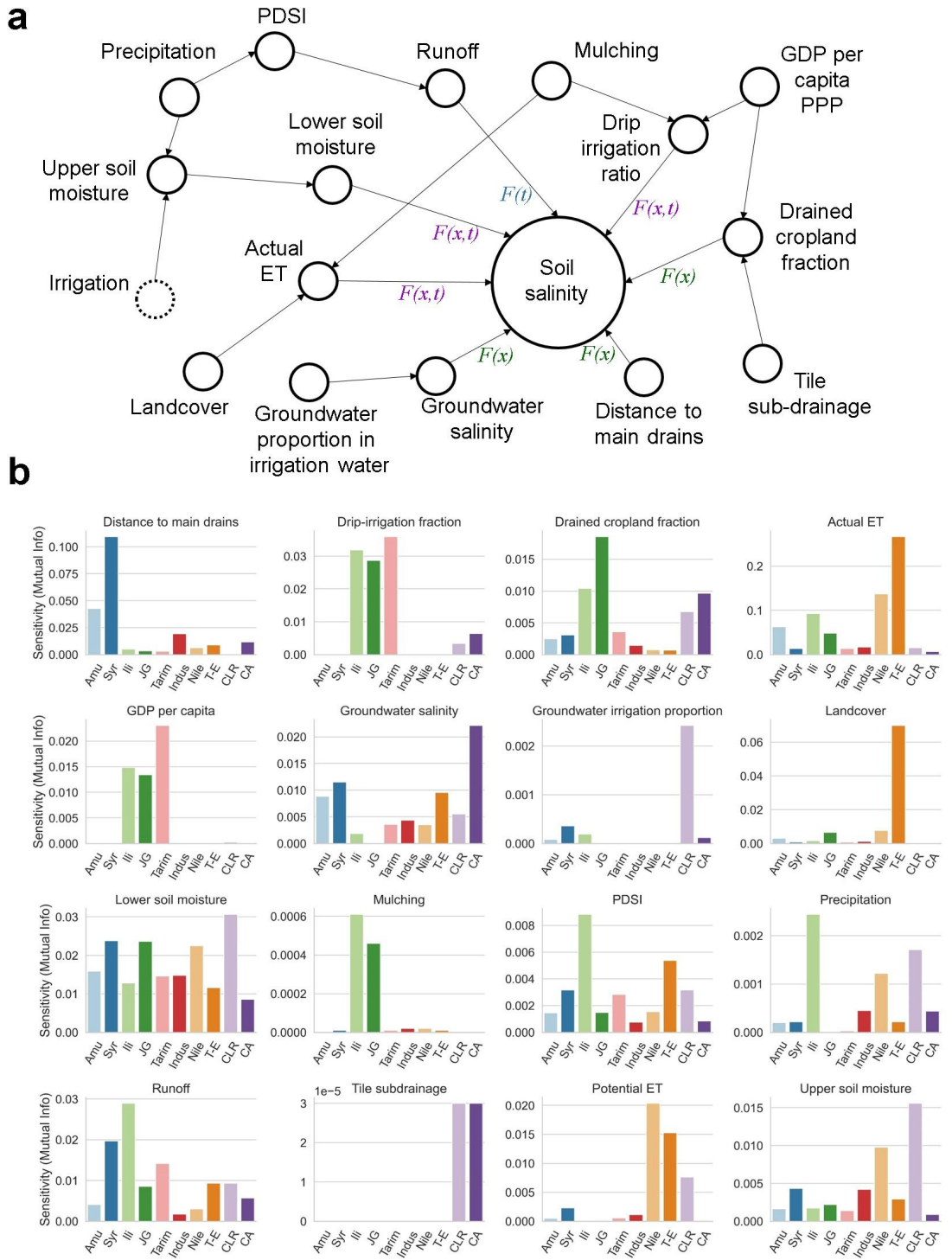

Fig. 5: **a**. Causal structure of the soil salinity Bayesian network. The directional arrows represent causal connections. F(x), F(t), and F(x, t) correspondingly represent the spatial, temporal, and spatial-temporal impacts of nodes on the soil salinity due to the data characteristics of the nodes. **b**. Comparison of the sensitivity patterns of the irrigation and water management-related nodes to the 'soil salinity' node of the Bayesian network. The sensitivity values are calculated based on the dependence of the entropy

reduction between the nodes of the Bayesian network. JG, Junggar. T-E, Tigris-Euphrates. CLR, Colorado. CA, Central Valley (San Joaquin). PDSI indicates Palmer drought severity index and ET indicates evapotranspiration.

**Opportunities for land degradation neutrality with reducing salinization**

Countries with different levels of development may have different prospects for reducing salinization and land degradation and for achieving LDN targets. Our analysis shows that different irrigation and drainage measures in basins with different water use and management practices have significant impacts on soil salinity. Shifts to water-saving, high-investment basins appear to reduce salinity and decrease soil moisture in the lower soil layer (Supplementary Fig. 12 and Supplementary Fig. 13). For example, in the Junggar Basin, from 2005 to 2015 the median soil salinity decreased by 8 dS/m as the fraction of drip irrigation increased from 0% to 49% (Supplementary Fig. 13). However, what improvements in irrigation and drainage are most economically feasible and cost-effective? The promotion of drip irrigation and subdrainage (tile or pipe subdrainage to remove salt accumulated by long-term drip irrigation below the root zone) to lower groundwater levels and thereby reduce topsoil salinity requires large-scale irrigation system reconstruction, which is challenging in low-income regions also with a lower percentage of government expenditure allocated to agriculture (Supplementary Fig. 11). In addition, drip irrigation and agricultural income appear to be positively correlated (Fig. 6). Drip irrigation can increase production and save water. As economic development increases, facilitated by increases in production and water savings, agriculture may shift to high-value fruit and forest industries. For example, increasing demand for wine, related to economic development, has led to planting patterns in the Central Valley and Junggar basins to gradually shift from cotton to grapes[20]. With increasing agricultural income, investments in improved irrigation systems may further increase especially in China and the United States with higher Agricultural orientation index (AOI) (Supplementary Fig. 11). The planting of forests and fruit can also increase soil organic matter and reduce salinity. Low-income river basins that mainly use flood irrigation still depend on conventional salinity leaching measures with low water efficiency. Improvements in irrigation and drainage technology also depend on industrial technology (e.g. to procure low-cost and efficient drip irrigation systems). Excessive extraction of groundwater contributes to the decline of groundwater levels, which reduces soil salinity (Fig. 6) but may also threaten the survival of natural vegetation, especially deep-rooted desert plants. Excessive extraction of groundwater can also reduce the sustainability of irrigation and increase energy consumption. In arid regions, lowering groundwater below 4-m depth can stop the contribution of groundwater to surface salt accumulation through rising capillary water[23]. Although groundwater depletion depends on the root depth of the vegetation[24], lowering the water table below 20-m depth can lead to serious groundwater depletion even for deep-rooted vegetation.

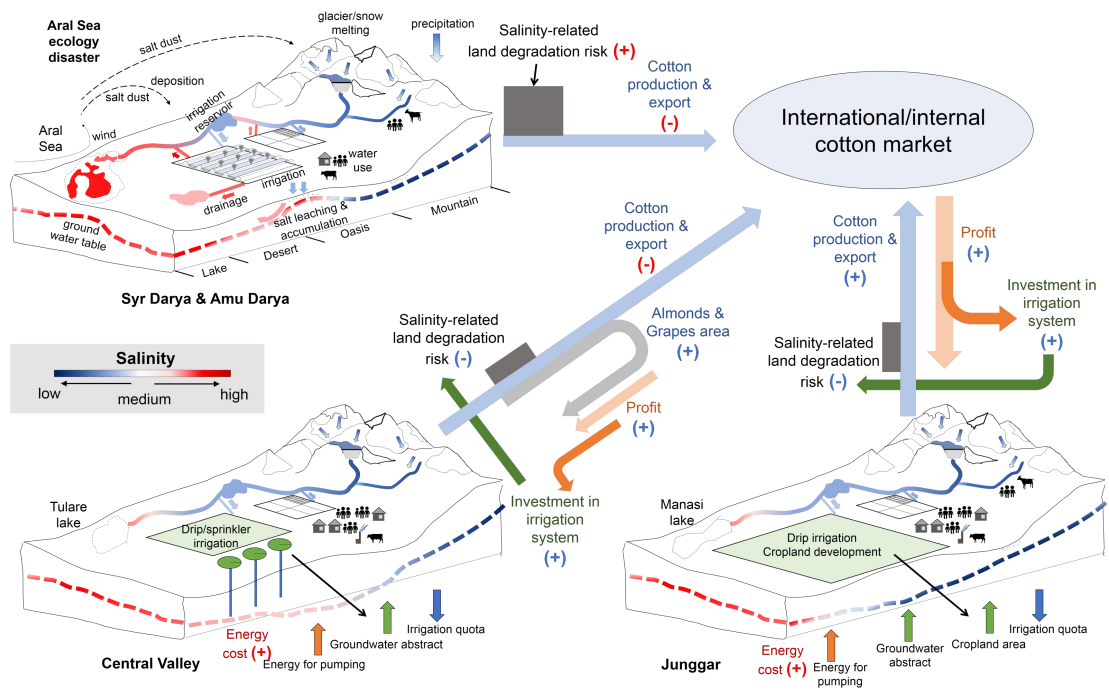

Fig. 6: Differences in the interactions of agricultural production, irrigation system investment, and the basin-scale salinity control in basins with similar mountain-oasis-desert-lake system and cotton planting. The light blue arrow indicates the value of cotton production (i.e., supply from the basin to the market), and the dark gray rectangle bordered by the light blue arrow indicates the risk of salinity-related land degradation due to cotton cultivation. The light gray arrow indicates the change in cropping structure, the light orange arrow indicates the total cropping profit, the dark orange arrow indicates the portion of the profit that is invested in irrigation system improvements, and the dark green arrow indicates the effect of the irrigation system improvements on the reduction of salinity-related land degradation risk.

## Discussion

Using remote sensing of salinity in major irrigated arid river basins around the world, we show that large-scale water management significantly impacts salinization, expanding on results from local and regional studies[5–7]. These local and regional studies use different data and approaches. Variability in approaches, data, and methods makes it difficult to align and compare the effects of water management on salinity. Our results help to fill a critical gap in our understanding of how water management and irrigation measures and interactions within the 'economy-irrigation system' influence salinization and our ability to achieve LDN.

The effectiveness of water management measures to control salinity varies at different stages of watershed development. After drip irrigation is widely used within a basin, drainage canals may be ineffective. Irrigated water can only infiltrate the zone below the roots, and farmers tend to use subdrainage. In watersheds with low levels of water management, uncertainties in irrigation and water management appear to be higher. Irrigation practices respond more passively to hydrological variations. For example, in the Amu Delta, farmers may block drainage to hold soil moisture in the root zone in dry years, while the volume of water applied for salt leaching (up to 6700 $m^3/ha$[25]) depends on upstream incoming water and farmers' subjective judgments. These factors may reduce the stability and robustness of long-term salinity control targets compared with the better-managed drip irrigation systems.

For LDN, recent studies have focused on the impacts of climate change, especially the effects of warming and droughts on desertification[26,27]. The problem of soil salinization, which is closely related to irrigation and water management, needs more focus in global efforts to achieve LDN. In many cases, anthropogenic factors, especially investment in irrigation systems, can be more important than climate change in salinization-related LDN issues. Irrigation system improvements can more quickly reduce salinization, increase yields, and provide higher income for farmers, compared to the complex international climate negotiations. Substantial improvements in irrigation systems may require deeper understanding and investments by governments and decision-makers because they require large initial investments and external financial support.

Proposals already exist to improve the irrigation system in the studied basins (e.g. recent proposals to promote drip irrigation in the Syr and Amu basins of Uzbekistan[28]). Local drip irrigation experiments have shown potential for improving water savings, salinity control, and agricultural yields. However, further study is necessary to understand the best management techniques in transboundary basins. For example, in the Syr and Amu basins, previous water conflicts have exacerbated the shrinking of the Aral Sea[12]. If drip irrigation is promoted, it remains uncertain whether the saved water will be used to reclaim croplands or to restore the Aral Sea. In addition, the promotion of drip irrigation in upstream regions may reduce the discharge of high-salinity return flow to the river (Fig. 6) and increase water available to downstream areas, thus improving water quality and quantity for downstream communities and agriculture. However, in cases like this, would upstream countries require downstream countries to compensate them for expenses associated with irrigation system improvements? Transboundary watershed and irrigation management may require win-win cooperation between countries to avoid conflicts[29].

Our results are based on spatio-temporal correlation analysis and causal machine learning, which are widely used in recent environmental studies[12,30]. BN provides more reliable explanations of causality and the relative contributions of the many factors related to salinization and watershed management, but the accuracy depends on a priori assumptions of causality and data uncertainty. Thus, the temporal and spatial scales of available data can affect our understanding of the causal relationships between watershed management and salinity. Unified use of datasets, however, facilitates the comparison of watersheds. Discretization of node parameters based on expert knowledge may also affect the sensitivity of salinity to different variables across basins. Due to the absence of related records in the multi-source soil samples data used, some factors (e.g., the number of subsamples per sample[3]) affecting salinity prediction accuracy were under-considered in this study, which may lead to uncertainty in the soil salinity prediction model. In the PCR-Globwb model, detailed irrigation methods may not be differentiated in the model. This may bring errors to the data (e.g. lower layer soil moisture) of watersheds with higher fractions of drip irrigation. Fortunately, due to the inclusion of actual irrigation area expansion data (such as in the Junggar basin), the low irrigation quota of drip irrigation may have been quantified partially and indirectly by the representation of deficit irrigation[22] in the model under the condition of extremely high water consumption ratios. In addition, in flood-irrigated regions, the difference in actual irrigation volume applied in the field is not perfectly simulated, possibly due to the lack of integration of the water used for salt leaching (Supplementary Table 3). Therefore, quantifying crop types, irrigation methods, and irrigation volumes at higher resolutions may improve future global hydrological model simulations, reduce the uncertainty of models, and add details to our understanding of local and regional impacts of water management on salinity. In addition, with BNs representing the causal relationship between salinity and water management, we can further predict the impacts of irrigation and water management on soil salinity under various future scenarios (e.g. climate, hydrology, land use), where the refinement of potential changes in irrigation and water management practices may be crucial.

In summary, we quantified the impacts of irrigation and water management on watershed-scale soil salinity in dry croplands. With population growth and climate changes, increasing food demand may increase the demand for irrigation[31], which is expected to lead to more severe dryland expansion[32] and higher land degradation risks. It also poses greater challenges to areas that traditionally use flood irrigation to control soil salinity, common in developing countries. Given these expected future changes, the results of this study emphasize the need to improve irrigation systems for soil salinity control and to take timely actions. The contribution of specific drivers of irrigation and water management to soil salinization highlights the need to incorporate these anthropogenic drivers into future global models that couple hydrology, agriculture, and economy. Our results provide direction for future salinity controls and further realization of LDN.

**Methods**

**Soil salinity estimation in Google Earth Engine from 1984 to 2018**

To establish a soil salinity estimation model, we combined Landsat satellite data and in situ soil salinity samples (Supplementary Fig. 5). We collected data from in-situ soil salinity samples (total of 2354 sites spanning 1984 to 2019) of the ten river basins from field sampling of our team (in Xinjiang, China, and Central Asia), WOSIS[33] database, and literature (Supplementary Fig. 3 and Supplementary Fig. 4). We normalized different sampling dates, depths, and salinity measurement methods, to the daily scale, the depth of 0-30 cm of topsoil, and the electrical conductivity of soil-saturated extracts. We identified samples located in cropland (filtered with the GFSAD30 cropland extent data for 2015, USGS) with the month of the sampling practice and accurate coordinate information and used these samples in our modeling. To match satellite image acquisition dates, which can affect prediction accuracy[3], we determined specific sampling dates.

For samples lacking exact dates, we assume the date of sampling is the 15th of the month. For soil samples at various depth layers, we weighed the salinity of multiple soil layers to a mean salinity value of the 0-30 cm layer. If the upper depth of the sample was deeper than 30 cm, it was excluded. For the normalization of different salinity measurement methods, we used empirical conversion coefficients[3] (Supplementary Table 2) depending on the soil texture and unified them (EC1:1, EC1:2, EC1:2.5, EC1:5, soil salt content) to ECe (electrical conductivity at the saturated extract phase). The samples of the Central Valley and Colorado basins in WOSIS are too dense compared to Central Asia and other regions; thus to balance the potential performance of our prediction model in each basin, we randomly sampled the sample set of croplands in the Central Valley and Colorado basins from WOSIS with the same sample density (the number of samples divided by the area of cropland) of the remaining sample set including other eight basins.

Subsequently, in the GEE platform, based on the temporal and spatial information of samples, we extracted their Landsat image bands and calculated various vegetation and soil salinity indices (Supplementary Table 1) used as the covariate predictor of a Random Forest model to estimate the large-scale salinity. With parameter adjustment (using 'smileRandomForest' function in GEE with Random Forest parameters set as numberOfTrees:300, variablesPerSplit:9, bagFraction:0.7, minLeafPopulation:2), R-squared of the model in the validation phase was 0.75 with 10% data subset used for validation (Supplementary Fig. 6), which can satisfy the accuracy requirement compared to the average performance of previous regional salinity prediction models[3] using Landsat data and machine learning. To make full use of the information of all samples despite the spectral differences of different Landsat satellites, we use empirical conversion coefficients[34] to harmonize Landsat 5 (used for samples from 1984 to 2011), Landsat 7 (for 2012-2013), and Landsat 8 (for 2014-2019). In addition, some grid soil property-related datasets with high resolutions available on the GEE platform are also included in the modeling (Supplementary Table 1). Finally, we predicted the topsoil salinity of croplands (masked with cropland extent in CGLS-LC100 data in 2019) at a 30-m scale in arid regions from 1984 to 2018 in the globe after the model validation (Supplementary Fig. 6 and Supplementary Fig. 7). For each pixel, the output annual salinity is represented by the median of the derived pixel-scale salinity values of the available Landsat images with cloud coverage less than 10% from May 1st to October 1st of that year. To reduce the data calculation requirements for subsequent analysis of the relations to irrigation and water use data, we resampled the output soil salinity map to 1-km resolution, and 20,000 random points (as a representative of all salinity grids by sampling) were generated with ArcGIS within the cropland extent (cropland extent in CGLS-LC100 data in 2019) of the ten selected basins and the salinity time series from 1984 to 2018 were extracted for each point.

**Water use related data**

The 5 arcmin PCR-Globwb model outputs

PCR-Globwb[22] is a global hydrological model that integrates hydrological cycle and water use information, and its new version has a standard monthly output high spatial resolution of 5 arcmin from a 58-year simulation (1958-2015)[22]. It has been widely used to assess global issues related to groundwater, surface water, and irrigation water use[35–38]. The selected outputs include precipitation, potential evapotranspiration, total actual evaporation, total groundwater abstraction, upper layer (the depth of 0 to 30 cm) soil moisture, lower layer (the depth of 30 to 150 cm) soil moisture, and groundwater recharge. The annual scale averaged from the monthly scale time-series data of these variables are extracted to each random salinity site based on the location and year and used for subsequent analysis.

Irrigation methods

At present, there is no uniform and reliable information on the distribution and changes of drip irrigation methods on a global scale. Therefore, we mainly quantified the significant spatial-temporal changes in irrigation methods (from flood irrigation to micro-irrigation) in California of the western United States (which includes the Central Valley Basin) and Xinjiang in China (which covers Junggar, Tarim, and upper Ili basins) since the 1990s (Supplementary Fig. 10). The drip-irrigated fractions of croplands in the other basins in this study are quite low and not considered. Drip irrigation fraction data of Xinjiang came from the statistical yearbook data from 2010 to 2018. We linearly interpolated data from 2005 to 2009 assuming the drip irrigation fraction in 2005 as 0. We collected data on drip irrigation in California from 2001 to 2010 from the literature[20,39] and we linearly interpolated the data from 1985 to 2000 (with the total drip irrigation fraction trend since the 1980s[40]) and 2011 to 2018 (with the average increasing rate of 2007 to 2010). To analyze the impacts of film-mulching, we used the data of mulching irrigation extent in 2015 derived from Landsat data but limited to Xinjiang, China[41].

Irrigation water source

To compare the impact of differences in irrigation water sources on soil salinity, we used irrigation water source data[42] that quantified the proportion of groundwater and surface water used.

### Artificial drainage conditions

At present, there is no reliable information on the distribution and changes in artificial drainage canals and their drainage efficiency on a global scale. According to the drainage canal maps in the literature, we vectorized the distribution of the main drainage canals (Supplementary Fig. 8) of each basin with ArcGIS. However, the difference in the drainage efficiency of these canals is uncertain. In addition, we used an earlier estimated global artificial drainage density map[43] in irrigated areas based on statistical data. For subdrainage, we used tile drainage distribution data, which was limited to the continental United States[21].

### Groundwater salinity

A global water salinity database[44] has been established recently. Unfortunately, measured data are limited in the regions that this study focuses on, such as Central Asia and the Middle East. Therefore, to quantify the heterogeneity of groundwater salinity on a large scale, we collected distribution maps of groundwater salinity in these watersheds from the literature (Supplementary Fig. 9).

### Actual evapotranspiration

Actual evapotranspiration data incorporating remote sensing information may be useful to quantify information related to irrigation and water management. We used actual evapotranspiration data constructed from the vegetation index based on AVHRR images from 1983 to 2013[45].

### Basin-scale drought index and runoff

Basin-scale drought and runoff affect the vertical transport of soil salt and the volume of available irrigation water. We extracted the PDSI drought index and runoff of each basin from Terraclimate[46] and calculated the normalized variation (z-score of the period from 1958 to 2018).

### Landcover

Because we aimed to estimate the salinity of cropland, we needed to know whether each sample was located in farmland. We used a high-resolution cropland extent data Global Food Security-Support Analysis Data at 30 m (GFSAD) to filter sampling points located in cropland. When exporting the salinity map, we used the 100-m resolution cropland data from CGLS-LC100[47] available on GEE to limit the exported extent. The mutual transfer of cultivated land and bare land in land cover change can affect soil salinity. We used land cover data from ESA from 1992 to 2018 to quantify the impacts of land reclamation and abandonment.

### Economic development

Changes in per capita GDP may be useful to explain the promotion of micro-irrigation and cropland management level and thus affect soil salinity. Therefore, we used a global-scale data[48] of the temporal and spatial distribution of GDP per capita.

## BN analysis

A BN[49] is a directed acyclic graph used to model causality. It comprises nodes representing discrete or continuous quantities and directed edges which do not form a directed cycle. This means there are no loops or cycles (no path that leads from a node, via other nodes, back to itself). We used the Netica software to build the BN. The nodes of a BN represent random variables ($X_1,..., X_n$) with their joint probability distribution calculated as:

$$P(X) = P(X_1, X_2, ..., X_n) = \prod_{i=1}^{n} P(X_i|pa(X_i)) \quad \#(1)$$

where $pa(X_i)$ represents the value of the parent node of the node $X_i$. With expert knowledge, the prior conditional probability table of BN can be given. Then, we used the expectation-maximization algorithm to incorporate observational data extracted from the included articles into the BN to update the conditional probability table. To evaluate the sensitivity of salinity node to other nodes, we used the sensitivity analysis[12,50] of the BN with mutual information (MI) representing the entropy reduction of the distribution of the child node caused by the status change of the parent node. The greater the MI value, the higher the sensitivity of the child node to the parent node, accompanied by stronger causality. The formula of MI is as follows:

$$MI = H(Q) - H(Q|F) = \sum_q \sum_f P(q, f) \log_2 \left(\frac{P(q,f)}{P(q)P(f)}\right) \quad \#(2)$$

where H represents the entropy, Q represents the target node, F represents the set of other nodes and q and f represent the status of Q and F.

**Acknowledgments**

This research was supported by the National Natural Science Foundation of China (grant no. U1803243), the Strategic Priority Research Programme of the Chinese Academy of Sciences (grant no. XDA20060302), the National Natural Science Foundation of China (grant no 41877012), and High-End Foreign Experts Project. We also appreciate the insightful comments from the reviewers to help to improve this manuscript.

**Contributions**

H.S and G.L initiated this research and were responsible for the integrity of the work as a whole. H.S and G.L performed formal analysis, calculations and drafted the manuscript. H.S, G.L, E.H.S, S.W, J.D, C.C, F.U.O, X.M, X.Y, W.Z, Y.W, B.F, M.X, Y.Z, and Q.L were responsible for the data collection and analysis. G.L, P.D.M, T.V.D.V, O.H, and X.C contributed to resources and financial support.

**Ethics declarations**

The authors declare no competing financial interests.

**Data availability**

The predicted scale soil salinity data of the croplands in the ten arid basins from 1984 to 2018 can be accessed at https://doi.org/10.5281/zenodo.5633645 and other data can be accessed by contacting the first author (shihaiyang16@mails.ucas.ac.cn) based on reasonable request.

**Code availability**

The GEE code used in this work can be accessed by contacting the first author

(shihaiyang16@mails.ucas.ac.cn) based on reasonable request.